\documentclass[12pt]{article}
\usepackage{amsfonts,amssymb,amsmath,fullpage,pst-node,rotating}
\usepackage{multirow}
\newcommand{\R}{{\mathbb R}}
\newcommand{\Z}{{\mathbb Z}}

\newcommand{\cB}{{\cal B}}

\newcommand{\rd}{{\rm d}}

\newcommand{\cC}{{\cal C}}

\newcommand{\cX}{{\cal X}}

\newcommand{\be}{{\bf e}}
\newcommand{\bu}{{\bf u}}
\newcommand{\bz}{{\bf z}}
\newcommand{\bv}{{\bf v}}

\newcommand{\bx}{{\bf x}}
\newcommand{\bt}{{\bf t}}

\newcommand{\bw}{{\bf w}}

\newcommand{\mbel}{{\eta}}
\newtheorem{definition}{Definition}
\newtheorem{theorem}{Theorem}

\newtheorem{remark}{Remark}

\begin{document}
\title{Behaviour of trajectories near a two-cycle heteroclinic network}

\author{Olga Podvigina\\
%\footnote{corresponding author, e-mail: olgap@mitp.ru}\\
Institute of Earthquake Prediction Theory\\
and Mathematical Geophysics\\
84/32 Profsoyuznaya St, 117997 Moscow, Russian Federation}

\maketitle

\begin{abstract}
We study behaviour of trajectories near a type Z heteroclinic network which is
a union of two cycles. Analytical and numerical studies indicate
that attractiveness of this network can be associated with various kinds of
dynamics in its vicinity, one or both of these cycle being fragmentarily
asymptotically stable, or both being completely unstable. In the latter
case trajectories can switch irregularly between the cycles, or they can
make a certain number of turns around one of them before switching
to the other one. Regular behaviour of trajectories near a heteroclinic
network can be described using the notion of an omnicycle, which we introduce
in this paper. In particular, we use it to prove that the network
can be fragmentarily asymptotically stable even if both cycles are completely
unstable.
\end{abstract}

%\bigskip\noindent
%{\bf We study behaviour of trajectories near a heteroclinic network, comprised
%of two cycles. The trajectories can be attracted to one of the cycles, or they
%can switch regularly or irregularly between them. This gives the simplest
%example known in the literature of a heteroclinic network featuring chaotic
%nearby trajectories. To describe regular switching we introduce the notions of
%an omnicycle and its trail-stability, and provide conditions for such stability.}

\section{Introduction}\label{sec_intro}

There is a number of reasons for chaotic behaviour of trajectories in a
vicinity of heteroclinic network. This can be due to the complexity of the
network as, e.g., when a network involves nodes which are chaotic sets
\cite{a03,d95} or is a union of periodic orbits connected by
an infinite number of heteroclinic and homoclinic orbits \cite{rla11}.
Chaotic behaviour can be caused by chaotic forcing, even when the
amplitude of the added noise is arbitrarily small \cite{am03}.
When eigenvalues of linearisations near some of the equilibria in the network
are complex, the return map near a heteroclinic cycle may possess a horseshoe,
thus resulting in emergence of a strange attractor \cite{acl06,lab}.
Chaos due to the presence of complex eigenvalues of linearisations was found
in \cite{mechanism} and \cite{rod}, where either a horseshoe does not
exist, or its existence has not been investigated.
Strange attractors may also exist in periodically-perturbed
dynamical systems with an attracting heteroclinic network \cite{rl21,lr20}.

Examples of more primitive heteroclinic networks comprised of a finite number
of equilibria and heteroclinic connections feature only real eigenvalues
of linearisation near equilibria; in their vicinity irregular switching
between connections takes place, indicating chaotic behaviour of nearby
trajectories \cite{ac10,pd05,pr21}. Complexity of a network can be measured
by the number of splitting nodes (for the definition see, e.g., \cite{AshCasLoh18}).
For the networks studied in \cite{ac10,pd05,pr21}, it is the number
of equilibria having more than one outgoing connections (six, nine and five,
respectively). In the present paper we consider such a primitive network
which has only a single splitting node. Our numerical simulations indicate
a possibility of irregular switching of nearby trajectories.

A heteroclinic cycle is a union of nodes and connecting trajectories,
where all nodes are required to be distinct. A heteroclinic network is a
connected union of heteroclinic cycles. In this paper we introduce the notion of
an omnicycle, an ordered sequence of nodes and heteroclinic connections, where
repeating nodes are allowed. Regarded as a subset of the
phase space, an omnicycle is either a heteroclinic cycle or a network, however
in the latter case the stability properties of these objects can be very different.
Omnicycles were implicitly employed in \cite{pd05} to describe the behaviour of trajectories
near a heteroclinic network, where existence of an attracting omnicycle was
indicated by so-called ``regular cycling''. However, the terminology introduced
{\it ibid} was closely related to the specific structure of the considered
network, while the definition introduced in the present paper is of a general type.

Omnicycles, similarly to heteroclinic cycles and network, do not exist in a
generic dynamical systems, because a connection between saddles can be destroyed
by a small perturbation. However, they may exist and be robust in a system
where some restrictions are imposed, e.g., symmetries \cite{km95a,km04},
natural constrains, as in population dynamics or evolutionary game theory
\cite{ac10,hs94,hs98} or due to prescribed patterns of interaction, as in a
system of coupled cells or oscillators \cite{bu00,as10}.

An invariant set of a dynamical system is called asymptotically stable if it
attracts all trajectories from its small neighbourhood.
A non-asymptotically stable set can be stable in a weaker sense
and attract a positive measure set of initial conditions from its small
neighdourhood. In \cite{op12} this property was called fragmentary asymptotic
stability. Conditions for asymptotic stability or fragmentary asymptotic
stability for certain types of heteroclinic cycles are known for a long time,
see, e.g., Podvigina \cite{op12} or \cite{pcl20} and references therein.
Typically, conditions for stability of a cycle are derived by following
nearby trajectories and constructing the respective transition maps.
Since it is irrelevant how many times each particular node or connection
is present in the sequence, the existing conditions for stability of
cycles can be straightforwardly employed for stability of omnicycles.
The study of stability of a heteroclinic network is more difficult, because
a trajectory can follow different sequences of connections along the network.

If a fragmentarily asymptotically stable (f.a.s.) cycle is a subset of a
heteroclinic network then the whole network also is f.a.s. For all networks in
$\R^4$ that were studied in literature the converse is also true: if a
heteroclinic network is f.a.s. then at least one of it subcycles is f.a.s.
\cite{br94,ks94,cl14,cl16,pc15,pc16,pl19}. The same property holds true
for some networks in higher-dimensional subspaces, namely stability of
a network implies stability of at least one of its subcycles \cite{ac10,pcl18}.
The authors of \cite{pd05,pr21} presented examples of f.a.s. networks, where
all subcycles are completely unstable. To prove stability the authors have
shown that certain omnicycles, that are subsets of the network, are f.a.s.
(To be more precise, in the proofs of \cite{pd05} a different technique was
employed, which in fact is similar to our theorem \ref{th_4}).

\begin{figure}
\hspace*{12mm}\includegraphics[width=18cm]{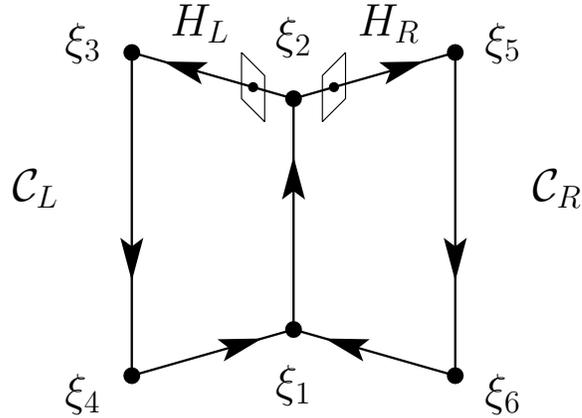}

{\Large
\vspace*{-53mm}\hspace*{72mm}$\xi_1$

\vspace*{-54mm}\hspace*{72mm}$\xi_2$

\vspace*{-7mm}\hspace*{44mm}$\xi_3$\hspace{51mm}$\xi_5$

\vspace*{39mm}\hspace*{44mm}$\xi_4$\hspace{51mm}$\xi_6$

\vspace*{-35mm}\hspace*{37mm}$\cC_L$\hspace*{63mm}$\cC_R$

\vspace*{-27mm}\hspace*{58mm}$H_L$\hspace*{17mm}$H_R$
}

\vspace{55mm}

\caption{The network studied in sections \ref{sec4} and \ref{sec5}. The
crossections $H_L$ and $H_R$ employed in the derivation of conditions of
stability are introduced in section \ref{sec3}.
\label{fig1}}
\end{figure}

In this paper we study stability of a network in $\R^6$ which is a union of
two cycles,
$\cX=\cC_L\cup\cC_R\subset\R^6$, admitted by a $\Z_2^6$-equivariant dynamical
system
\begin{equation}\label{eq_ode6}
\dot{\bf x}=f({\bf x}),\quad f:\R^6\to\R^6,
\end{equation}
where $\Z_2^6$ is generated by symmetries reversing the sign of one of the
spatial coordinates. Six equilibria involved in the network belong to
coordinate axes and heteroclinic connection belongs to coordinate
planes. The cycles are:
$$\cC_L:\,(\xi_1\to\xi_2\to\xi_3\to\xi_4\to)\hbox{ and }
\cC_R:\,(\xi_1\to\xi_2\to\xi_5\to\xi_6\to).$$

In terminology of \cite{op12}, $\cX$ is a {\it type Z} heteroclinic network
and $\cC_L$ and $\cC_R$ are {\it type Z} heteroclinic cycles. Generalisation
of this classification to omnicycles implies that any omnicycle, which is a
subset of $\cX$, is a {\it type Z} omnicycle.
Theorems in \cite{op12} that give necessary and sufficient
conditions for a type Z heteroclinic cycle to be f.a.s. are also applicable to
study stability of type Z heteroclinic omnicycles.

The conditions are inequalities that involve eigenvalues and eigenvectors of
transition matrices, whose entries depend on eigenvalues of Jacobians
$\rd f(\xi_j)$. For $\cC_L$ and $\cC_R$ the respective $4\times4$ transitions
matrices are sparse, hence allowing to calculate eigenvalues and eigenvectors.
Transition matrix of the omnicycle $\cC_{LR}$, where
$$\cC_{LR}:\,
(\xi_1\to\xi_2\to\xi_3\to\xi_4\to\xi_1\to\xi_2\to\xi_5\to\xi_6\to),$$
is full, therefore eigenvalues and eigenvectors can not be found analytically
in a general case. In order to be able to calculate them, we assume that some
eigenvalues of $\rd f(\xi_j)$ are much larger then the
others. For vanishing small eigenvalues we calculate eigenvalues
and eigenvectors of the respective sparse transition matrix. Because of
continuous dependence of eigenvalues and eigenvectors of a matrix
on its entries, the stability or instability conditions
are satisfied true for small
non-vanishing eigenvalues of $\rd f(\xi_j)$ as well.

%Ways to construct dynamical systems possessing certain robust heteroclinic
%cycles or networks are known for a long time. Moreover, the construction can
%be designed so that $\rd f(\xi_j)$ has prescribed eigenvalues. Therefore,
%based on the results we can prescribe eigenvalues in such a way that
%$\cC_L$ and $\cC_R$ are unstable and $\cC_{LR}$ is stable.
%We present numerical simulations supporting our analytical findings about
%stability or instability of cycles or omnicycles. While all they are unstable,
%trajectories can follow more complex sequences of equilibria, regularly
%or apparently irregularly and chaotic.

The paper is organised as follows: In Section \ref{sec2} we recall basic
definitions of robust heteroclinic cycles, networks and asymptotic stability
and in Section \ref{sec3} we discuss asymptotic stability of omnicycles.
In sections \ref{sec4} and \ref{sec5} we study analytically and numerically
asymptotic stability of network shown in fig.~\ref{fig1}. Namely, in section
\ref{sec4} we derive conditions of stability of the cycles $\cC_L$ and $\cC_R$
and show that the network can be f.a.s. while both cycles are completely
unstable. In section \ref{sec5} we present results of numerical simulations
of solutions of system (\ref{eq_ode6}), in particular, in the cases when
all $\cC_L$, $\cC_R$ and $\cC_{LR}$ are completely unstable.
By varying the r.h.s. of (\ref{eq_ode6}) we obtain examples of f.a.s.
omnicycles, different from $\cC_{LR}$, or of apparently chaotic
switching between $\cC_L$ and $\cC_R$.

\section{Background and definitions}\label{sec2}

In this section we introduce the notions of omnicycles and trail-stability,
and recall definitions that will be used in the following sections.

\subsection{Robust heteroclinic cycles, omnicycles and networks}\label{defhet}

Consider a smooth $\Gamma$-equivariant dynamical systems
\begin{equation}\label{eqsys}
\dot x=f(x),\quad f(\gamma {\bf x})=\gamma f({\bf x}),\quad\mbox{ for all }
\gamma\in\Gamma\hbox{ and }x\in\R^n,
\end{equation}
where we assume that $\Gamma\subset{\bf O}(n)$ is finite.
For a group $\Gamma$ acting on $\R^n$ the {\it isotropy group} of
the point $x\in\R^n$ is the subgroup
$\Sigma_x=\{\gamma\in\Gamma\ ~:~\ \gamma x=x\}$,
and a {\it fixed-point subspace} of a subgroup $\Sigma\subset\Gamma$
is the subspace
${\rm Fix}(\Sigma)=\{{\bf x}\in\R^n\ ~: \ \sigma {\bf x}={\bf x}\mbox{ for all } \sigma\in\Sigma\}$.
(For more details about equivariant dynamical systems see, e.g.,
Golubitsky and Stewart \cite{GSS}.)

Let $\xi_1,\ldots,\xi_m$ be hyperbolic equilibria of (\ref{eqsys}),
where $\xi_j\ne\xi_k$ for any $1\le j,k\le m$, $j\ne k$, and
$\kappa_{j,j+1}\ne\varnothing$ be a set of trajectories from $\xi_j$ to
$\xi_{j+1}$, where $\xi_{m+1}=\xi_1$ is assumed.
A {\it heteroclinic cycle} is an invariant set $X\subset\R^n$ which is a union
of equilibria $\{\xi_1,\ldots,\xi_m\}$ and
heteroclinic connections $\{\kappa_{12},\ldots,\kappa_{m1}\}$.
A {\it heteroclinic network} is a connected union of a finite
number of heteroclinic cycles.

\begin{definition}\label{def01}
An {\it omnicycle} is a union of equilibria $\{\xi_1,\ldots,\xi_m\}$ and
heteroclinic connections, $\{\kappa_{12},\ldots,\kappa_{m1}\}$, where
$\kappa_{ij}\subset W^u(\xi_i)\cap W^s(\xi_j)$ and $\kappa_{ij}\ne\varnothing$.
\end{definition}
I.e., the difference between a heteroclinic cycle and an omnicycle is that
for the omnicycle repeating equilibria are allowed.

By analogy with a building block of a heteroclinic cycle \cite{pc15}, we
introduce building block of an omnicycle.
\begin{definition}\label{bb}
Given an omnicycle $\cX=\{\xi_1,\ldots,\xi_m;\kappa_{12},\ldots,\kappa_{m1}\}$,
we say that its subset
$\{\xi_1,\ldots,\xi_j;\kappa_{12},\ldots,\kappa_{j,j+1}\}$ and the symmetry
$\gamma\in\Gamma$ is a {\it building block} of $\cX$ if
$\gamma\xi_i=\xi_{i+j}$ and $\gamma\kappa_{i,i+1}=\kappa_{i+j,i+j+1}$ for
any $\xi_i\in\cX$ and $\kappa_{i,i+1}\subset\cX$.
\end{definition}

A heteroclinic cycle (or an omnicycle, or a network) $\cX$ is called
{\it robust}, if any connection $\kappa_{ij}\subset\cX$ belongs to a flow-invariant
subspace $P_{ij}$. In system (\ref{eqsys}) the invariance of $P_{ij}$
typically follows from its equivariance, namely that
$P_{ij}={\rm Fix}(\Sigma_{ij})$
for some $\Sigma_{ij}\subset\Gamma$. By $\Delta_j$ we denote the isotropy
subgroup of $\xi_j$ and $L_j={\rm Fix}(\Delta_j)$.
Denote by $P_{ij}^{\perp}$ and by $L_j^{\perp}$ the orthogonal complement to
$P_{ij}$ and $L_j$, respectively, in $\R^n$.

Given an equilibrium $\xi_j\in L_j=P_{j-1,j}\cap P_{j,j+1}$ that belongs to a
robust heteroclinic cycle $\cC=(\xi_1\to\xi_2\to...\to\xi_m\to)$, the
eigenvalues of the Jacobian $\rd f(\xi_j)$ can be divided into {\it radial}
(the associated eigenvectors belong to $L_j$), {\it contracting} (the
eigenvectors belong to $P_{j-1,j}\ominus L_j$), {\it expanding} (the
eigenvectors belong to $P_{j,j+1}\ominus L_j$) and {\it transverse} (the remaining
ones), where $P\ominus L$ denotes the orthogonal complement to $L$ in $P$.

\begin{definition}\label{def02} (Adapted from \cite{op12})
A heteroclinic network $\cX$ (cycle or omnicycle) is called of {\it type Z} if\\
$\bullet$ for any $\xi_j\in\cX$ we have that $\dim P_{i,j}=\dim P_{j,k}$ for
any incoming connection $\kappa_{i,j}$ and any outgoing connection $\kappa_{j,k}$;\\
$\bullet$ for any $\kappa_{i,j}\subset\cX$ the isotropy subgroup $\Sigma_{i,j}$
decomposes $P_{i,j}^{\perp}$ into one-dimensional isotypic components.
\end{definition}

%An equilibrium in a network $\cX$ might belong to several heteroclinic cycles.
%In such a case the radial subspace is the same for all cycles, by contracting
%eigenvectors (or eigenvalues) we denote the ones that are contracting for
%at least one subcycle of $\cX$ and similarly for expanding ones. The rest
%of eigenvectors (or eigenvalues) are called transverse. To avoid ambiguity,
%whenever necessary we write $\cC$-contracting ($\cC$-expanding or
%$\cC$-transverse) eigenvectors or eigenvalues to denote objects related to
%the cycle $\cC$ and use $\cX$-contracting ($\cX$-expanding or
%$\cX$-transverse) for the ones related to $\cX$.

\subsection{Stability}\label{sec_stab}

Denote by $\Phi({\bf x},\tau)$ a trajectory of system (\ref{eqsys}) through
the point ${\bf x}\in\R^n$ and
by $\Phi({\bf x},(\tau_1,\tau_2))=
\cup_{\tau\in(\tau_1,\tau_2)}\Phi({\bf x},\tau)$ the subset
of this trajectory corresponding to the time interval $(\tau_1,\tau_2)$.
For a set $X\subset\R^n$ and $\varepsilon>0$, the
$\varepsilon$-neighbourhood of $X$ is
\begin{equation}\label{ep_nei}
B_{\varepsilon}(X)=\{{\bf x}\in R^n:\ d({\bf x},X)<\varepsilon\}.
\end{equation}
Given a compact invariant set $X$ of (\ref{eqsys}), its
$\delta$-local basin of attraction defined as
\begin{equation}\label{del_bas}
\cB_{\delta}(X)=\{{\bf x}\in\R^n:\ d(\Phi({\bf x},\tau),X)<\delta\hbox{ for any }
\tau\ge0\hbox{ and }\lim_{\tau\to\infty}d(\Phi({\bf x},\tau),X)=0\}.
\end{equation}

\begin{definition}\label{def1}
An invariant set $X$ is called {\it asymptotically stable}, if for any
$\delta>0$ there exists an $\varepsilon>0$ such that
$$B_{\varepsilon}(X)\subset\cB_{\delta}(X).$$
\end{definition}

\begin{definition}\label{def2}
An invariant set $X$ is called {\it fragmentarily asymptotically
stable}, if for any $\delta>0$
$$\mu(\cB_{\delta}(X))>0.$$
(Here $\mu$ is the Lebesgue measure of a set in $\R^n$.)
\end{definition}

\begin{definition}\label{def3}
An invariant set $X$ is called {\it completely unstable}, if there exists
$\delta>0$ such that $\mu(\cB_{\delta}(X))=0$.
\end{definition}

\begin{definition}\label{defn1}
Given an omnicycle $\cX=\{\xi_1,\ldots,\xi_m;\kappa_{12},\ldots,\kappa_{m1}\}$,
we say that a trajectory $\Phi({\bx},(\tau^{in},\tau^{fin}))$
$\varepsilon$-{\it follows} this omnicycle if
$$\Phi({\bx},(\tau^{in},\tau^{fin}))\subset N_{\varepsilon}(\cX)$$
and there exist a number $1\le j_0\le m$ and time instances
$\{\tau_0',\tau_0'',\tau_1',\tau_1'',...,\tau_S',\tau_S''\}$, where
$\tau^{in}\le\tau_0'<\tau_0''<\tau_1'<...<\tau_S'<\tau_S''\le\tau^{fin}$,
such that
$$
\renewcommand{\arraystretch}{1.5}
\begin{array}{l}
\Phi({\bx},(\tau_s',\tau_s''))\subset B_{\varepsilon}(\xi_j),
\hbox{ where }j=j_0+s({\rm mod}\, m),\hbox{ for all }0\le s\le S,\hbox{ and}\\
\biggl((\cup_{0\le s<S}\Phi({\bx},(\tau_s'',\tau_{s+1}'))\cup
\Phi({\bx},(\tau^{in},\tau_0))\cup
\Phi({\bx},(\tau_S,\tau^{fin}))\biggr)\bigcap
\biggl(\cup_{1\le j\le m}B_{\varepsilon}(\xi_j)\biggr)=\varnothing.
\end{array}
$$
\end{definition}

\begin{definition}\label{defn2}
Given an omnicycle $\cX$ of (\ref{eqsys}), its
$\delta$-local basin of attraction is defined as
\begin{equation}\label{del_omni}
\cB^{\rm omni}_{\delta}(\cX)=
\{{\bf x}\in\R^n:\ \Phi({\bf x},(0,\tau))\ \delta-\hbox{follows }\cX
\hbox{ for any }
\tau>0\hbox{ and }\lim_{\tau\to\infty}d(\Phi({\bf x},\tau),\cX)=0\}.
\end{equation}
\end{definition}

\begin{definition}\label{defn3}
An omnicycle $\cX$ is called {\it trail-stable}, if for any $\delta>0$
$$\mu(\cB^{\rm omni}_{\delta}(\cX))>0.$$
\end{definition}

\begin{remark}\label{rem00}
For an omnicycle which is a heteroclinic cycle the notions of
f.a.s. and trail-stability are equivalent, because trajectories near the cycle
visit the equilibria in the prescribed order specified in the definition of
the cycle. An omnicycle, which is not a cycle, can be f.a.s. according to
definition \ref{def2} without being trail-stable.
\end{remark}

\begin{remark}\label{rem01}
Suppose $\cX$ is a robust type Z omnicycle, which is not a heteroclinic cycle,
in a $\Gamma$-equivariant system. Then $\cX$ is not asymptotically stable.
This can be shown by arguments similar to the ones employed in the proof
of theorem 3.1 in \cite{pcl18}.
\end{remark}

\section{Stability of omnicycles}\label{sec3}

In this section we state conditions for trail-stability of type Z omnicycles.
Due to similarity between heteroclinic cycles and omnicycles, the theorem
of \cite{op12}, which gives conditions for fragmentary asymptotic stability of
heteroclinic cycles of type Z, is also applicable to omnicycles of that type.
The conditions involve eigenvalues and eigenvectors of transition matrices,
which are constructed from local and global maps defined below.

\subsection{Local and global maps}
\label{locstr}

To study stability of an omnicycle, similarly to heteroclinic cycles
and networks, the behaviour
of nearby trajectories is approximated by local and global maps. By
$N_{\tilde\delta}(\xi_j)$ we denote a neighbourhood of $\xi_j$ bounded by
$|a_s|=\tilde\delta$, where $a_s$ are coordinates in the basis comprised
of eigenvectors of $\rd f(\xi_j)$. We assume $\tilde\delta$ to be fixed and
small, so that in $N_{\tilde\delta}(\xi_j)$ the equation (\ref{eqsys})
can be approximated by its linearisation $\rd f(\xi_j)$, and
$\tilde\delta\gg\varepsilon$ and $\delta$, where $\varepsilon$ and $\delta$ are
those employed in the definitions of stability in subsection \ref{sec_stab}

Given an omnicycle $\cC:\,(\xi_1\to...\to\xi_m\to)$, let
$(\bu_j,v_j,w_j,{\bf z}_j)$ be local coordinates
near $\xi_j$ in the basis, where radial eigenvectors come the first
(the respective coordinates are $\bu_j$), followed by the contracting
and the expanding eigenvectors, the transverse eigenvectors being the last.
(If repeating equilibria are present in the sequence, i.e. $\xi_i=\xi_j$
for $i\ne j$, then such decompositions for eigenvectors of $\rd f(\xi_i)$
and $\rd f(\xi_j)$ can be different.)
Denote by $H^{(in)}_{j-1,j}$ the face of $N_{\tilde\delta}(\xi_j)$ that
intersects with the connection $\kappa_{j-1,j}$ and
by $H^{(out)}_{j,j+1}$ the face intersected by $\kappa_{j,j+1}$.
In literature, $H^{(in)}_{j-1,j}$ and $H^{(out)}_{j,j+1}$ are often
called crossections or Poincar\'e sections. For a trajectory close to the
cycle the local map relates coordinates of its intersection with
$H^{(in)}_{j-1,j}$ to the coordinates of its intersection with
$H^{(out)}_{j,j+1}$.

Let the superscripts $^{in}$ and $^{out}$ denote the coordinate in
$H^{(in)}_{j-1,j}$ and $H^{(out)}_{j,j+1}$, respectively.
The local map $\phi_j:H^{(in)}_{j-1,j}\to H^{(out)}_{j,j+1}$ (see, e.g.,
\cite{km95a} or \cite{op12}) is
\begin{equation}\label{mapl0}
v^{out}=K_{1j}(w^{in})^{-c_j/e_j}\hbox{ and }
\bz^{out}={\bf K}_j\bz^{in}(w^{in})^{-\bt_j/e_j},
\end{equation}
where $K_{1j}$ and ${\bf K}_j$ are constants (for a fixed $\tilde\delta$), and
$c_j,e_j,\bt_j$ are the contracting, expanding and transverse eigenvalues of
$\rd f(\xi_j)$, respectively.
The coordinates $\bu^{in},v^{in},\bu^{out}$ and $w^{out}$ are ignored, because
they are irrelevant in the study of stability.

Near the connection $\kappa_{j,j+1}$ the system (\ref{eqsys}) can be
approximated by a global map (also called a {\em connecting diffeomorphism})
$\psi_{j,j+1}:H^{(out)}_{j,j+1}\to H^{(in)}_{j,j+1}$,
which relates coordinates of the
point where a trajectory exits $N_{\tilde\delta}(\xi_j)$ to coordinates of
the point where it enters $N_{\tilde\delta}(\xi_{j+1})$.
The global map is predominantly linear
\begin{equation}\label{mapgl}
(w^{j+1,in},\bz^{j+1,in})=A'_j(v^{j,out},\bz^{j,out}),
\end{equation}
where $A'_j$ is an $(n_t+1)\times(n_t+1)$ matrix. For type Z omnicycles
the matrix $A'_j$ is a product of a diagonal matrix $A''_j$ and
a permutation matrix $A_j$.

Denote by $g_j$ the superpositions of the local and global maps:
$$g_{j}=
\phi_{j}\circ\psi_{j-1,j}:\, H^{(out)}_j\to H^{(out)}_{j+1}.$$
We call the set of maps $\{g_{1:m}\}=\{g_{1},\ldots,g_{m}\}$,
$g_{j}:\R^{n_t+1}\to\R^{n_t+1}$,
{\it the collection of maps associated with the omnicycle}
$\cC=\{\xi_1,\ldots,\xi_m\}$.
For a given a collection of maps $\{g_1^m\}$ we define
superpositions
\begin{equation}\label{mapg}
G_j=g_{j-1}\circ\ldots\circ g_{1}\circ g_{m}\circ\ldots
\circ g_{j+1}\circ g_{j},
\quad G_j:\,H^{(out)}_j\to H^{(out)}_j,
\end{equation}
which approximate the behaviour of trajectories near the cycle and are
called {\it return maps}.

\subsection{Transition maps and transition matrices}
\label{trmatr}

Let us introduce new coordinates:
\begin{equation}\label{newc}
\mbel=(\ln|w|,\ln|z_1|,...,\ln|z_{n_t}|).
\end{equation}
Due to the smallness of $\varepsilon$, in these coordinates
the maps $g_{j}$ become approximately linear, $\mbel^{j+1}=M_j\mbel^j$,
where the transition matrix $M_j$ is a product of permutation matrix
$A_j$ that relates local coordinates near $\xi_{j+1}$ to the ones near $\xi_j$
and the matrix
\begin{equation}\label{esm}
B_j:=\left(
\begin{array}{ccccc}
b_{j,1}&0&0&\ldots&0\\
b_{j,2}&1&0&\ldots&0\\
b_{j,3}&0&1&\ldots&0\\
.&.&.&\ldots&.\\
b_{j,N}&0&0&\ldots&0
\end{array}
\right).
\end{equation}
The entries
$b_{j,l}$ of the matrix $B_j$ depend on the eigenvalues of the
linearisation $df(\xi_j)$ of (\ref{eqsys}) near $\xi_j$ as follows
\begin{equation}\label{coeB}
b_{j,1}=c_j/e_j\mbox{ and }b_{j,l+1}=-t_{j,l}/e_j,\ 1\le l\le n_t,\ 1\le j\le m.
\end{equation}
Transition matrices of the superposition $G_j$ defined in (\ref{mapg})
are the products\\
 ${\cal M}^{(j)}=M_{j-1}\ldots M_1M_m\ldots M_{j+1}M_j$.
%Denote by $\lambda_s$ the eigenvalues
%of matrices ${\cal M}^{(j)}$ (they are independent of $j$, since all matrices
%${\cal M}^{(j)}$ are similar) enumerated in the descending order of their real parts
%(generically all the real parts are distinct except for pairs of complex
%conjugate eigenvalues). Let also ${\bf w}^{j,s}$ denote the eigenvector
%of the matrix ${\cal M}^{(j)}$ associated with the eigenvalue $\lambda_s$.

Consider the matrix $M:={\cal M}^{(1)}=M_m\ldots M_1:\R^N\to\R^N$; it is
a product of the basic transition matrices of the form (\ref{esm}). We separate
the coordinate vectors ${\bf e}_l$, $1\le l\le N$, into two groups.
The first group is comprised of the vectors ${\bf e}_l$ for which there exist
such $k$ and $j$ that $(A^{(j)})^kA_{j-1,1}{\bf e}_l={\bf e}_1$ (recall that
$A_j$ are permutation matrices), the second one incorporates the remaining
vectors. Denote by $V^{\rm sig}$ and $V^{\rm ins}$ the subspaces spanned
by vectors from the first and second group, respectively (the superscripts
``ins'' and ``sig'' stand for significant and insignificant). In the basis where
significant vectors come the first, matrix $M$ takes the form
\begin{equation}\label{mM}
M=\left(
\begin{array}{c|c}
\\
M^{\rm sig} & 0 \\
\\
\hline
M^{\rm ins} &
\begin{array}{cccc}
1&0&\ldots&0\\
0&1&\ldots&0\\
.&.&\ldots&.\\
0&0&\ldots&1
\end{array}
\end{array}
\right).
\end{equation}

\begin{theorem}\label{th_22} (Adapted from \cite{op12}.)
Let $V^{\rm sig}$ and $V^{\rm ins}$ be the subspaces defined above.
\begin{itemize}
\item[(a)] The subspace $V^{\rm ins}$ is $M$-invariant and
all eigenvalues associated with the eigenvectors from this subspace are one.
\item[(b)] Generically all components of eigenvectors that do not
belong to $V^{\rm ins}$ are non-zero.
\end{itemize}
\end{theorem}

\begin{remark}\label{rem31}
In the study of stability instead of whole heteroclinic cycle or omnicycle one
can consider its building block, see remark \ref{bb}. In such a case the
transition matrix take the form $M^{\rm block}=AM$, where $A$ is the permutation
matrix relating bases $\{{\bf e}_l\}$ and $\{\gamma{\bf e}_l\}$ and
$M$ has the form (\ref{mM}). The statement (a) therefore becomes \cite{op12}
``The subspace $V^{\rm ins}$ is $M$-invariant and the absolute value
of all eigenvalues associated with the eigenvectors from this subspace is one.''
\end{remark}

We call {\it insignificant} the eigenvalues associated with eigenvectors
from $V^{\rm ins}$, and {\it significant} the rest ones. Generically
the absolute values of all significant eigenvalues differ from one.
Below $\lambda_{\max}\ne 1$ denotes the largest significant eigenvalue of
a transition matrix ${\cal M}^{(j)}$.

The construction of transition matrices is identical for heteroclinic cycles
and omnicycles. Whether or not a set of trajectories stays close to a
given sequence of connections is independent of presence or absence of
repeating equilibria or connections in this sequence. Therefore the
theorem of \cite{op12} that gives necessary and sufficient conditions
for a heteroclinic cycle to be f.a.s. can be extended to an omnicycle
as follows.

\begin{theorem}\label{th_4}(Adapted from \cite{op12}.)
Let $M_j$ be basic transition matrices of a collection of maps $\{g_1^m\}$
associated with an omnicycle of type Z.
(For type Z omnicycles the matrices are of the form (\ref{esm}).)
Denote by $j=j_1,\ldots j_L$ the indices, for which $M_j$ involves negative
entries; all entries are non-negative for all remaining $j$. Denote
by $\lambda_{\max}$ the dominant eigenvalue of ${\cal M}^{(j)}$ and
by $\bw_{\max}$ the associated eigenvector.
\begin{itemize}
\item[(a)] If for at least one $j=j_l+1$
the matrix ${\cal M}^{(j)}$ does not satisfy the following conditions
\begin{itemize}
\item[(i)]$\lambda_{\max}$  is real;
\item[(ii)] $\lambda_{\max}>1$;
\item[(iii)] $w_l^{\max}w_q^{\max}>0$ for all $l$ and $q$, $1\le l,q\le N$.
\end{itemize}
then the omnicycle is trail-unstable.
\item[(b)] If the matrices ${\cal M}^{(j)}$ satisfy the above conditions (i)-(iii)
for all $j$ such that $j=j_l+1$, then the omnicycle is trail-stable.
\end{itemize}
\end{theorem}

\section{Fragmentary asymptotic stability: analytical results}\label{sec4}

In this section we apply theorem \ref{th_4} to study dynamics near the network
$\cX=\cC_L\cup\cC_R$ shown in fig.~\ref{fig1}. The network exists in a
$\Gamma$-equivariant system (\ref{eqsys}) in $\R^6$, where the group
$\Gamma\cong\Z^6_2$ is generated by symmetries inverting the sign of one
of coordinates. The equilibria in the network belong to
coordinate axes, the connections are one-dimensional and belong to coordinate
planes, and the equilibria are stable in directions transverse to the network.

According to theorem \ref{th_4}, stability of a cycle or an omnicycle depends on
the dominant eigenvalue and the associated eigenvectors of the respective
transition matrices. Transition matrices of the cycles $\cC_L$ and $\cC_R$ have
two-dimensional significant subspace, therefore we can calculate the dominant
eigenvalues and the associated eigenvectors.

The significant subspace of the transition matrix of the omnicycle
$$\cC_{LR}=(\xi_1\to\xi_2\to\xi_3\to\xi_4\to\xi_1\to\xi_2\to\xi_5\to\xi_6\to)$$
has dimension four, therefore in a general case its eigenvalues can not be found
analytically. To show that the omnicycle can be trail-stable, we calculate
the dominant eigenvalue and the associated eigenvector in the case when
some eigenvalues of $\rd f(\xi_j)$ vanish. Since the conditions for
stability are non-strict inequalities, the inequalities remains true
for small non-vanishing eigenvalues as well. Under our assumption that the chosen
eigenvalues of $\rd f(\xi_j)$ vanish, the conditions for stability of $\cC_{LR}$
are not compatible with conditions for stability of $\cC_L$ and $\cC_R$. So,
we prove that the network $\cX=\cC_L\cup\cC_R$ can be f.a.s. while both
subcycles are completely unstable.

\subsection{Stability of cycles $\cC_L$ and $\cC_R$}
\label{sec31}

Since the system under consideration is $\Z_2^6$-symmetric, the eigenvalues
of $\rd(\xi_j)$ coincide with the Cartesian basis vectors $\be_j$, $1\le j\le6$.
In crossections to the connections comprising the cycle $\cC_L$ we take
the following local bases:
\begin{equation}\label{bases}
H^{out}_{12}:\{\be_3,\be_4,\be_5,\be_6\},\
H^{out}_{23}:\{\be_4,\be_1,\be_5,\be_6\},\
H^{out}_{34}:\{\be_1,\be_2,\be_5,\be_6\},\
H^{out}_{41}:\{\be_2,\be_3,\be_5,\be_6\}.
\end{equation}
In agreement with (\ref{esm}) the basic transition matrices are:
\begin{equation}\label{m14}
\renewcommand{\arraystretch}{1.5}
\begin{array}{cc}
M_{412}=\left(
\begin{array}{cccc}
b_{132}&1&0&0\\
b_{142}&0&0&0\\
b_{152}&0&1&0\\
b_{162}&0&0&1
\end{array}
\right),
&
M_{123}=\left(
\begin{array}{cccc}
b_{243}&1&0&0\\
b_{213}&0&0&0\\
b_{253}&0&1&0\\
b_{263}&0&0&1
\end{array}
\right)\\
M_{234}=\left(
\begin{array}{cccc}
b_{314}&1&0&0\\
b_{324}&0&0&0\\
b_{354}&0&1&0\\
b_{364}&0&0&1
\end{array}
\right),
&
M_{341}=\left(
\begin{array}{cccc}
b_{421}&1&0&0\\
b_{431}&0&0&0\\
b_{451}&0&1&0\\
b_{461}&0&0&1
\end{array}
\right),
\end{array}
\end{equation}
where
\begin{equation}\label{bbr}
b_{ijk}=-\mu_{ij}/\mu_{ik}
\end{equation}
and $\mu_{ij}$ is the eigenvalue of $\rd f(\xi_i)$ associated with the
eigenvector $\be_j$.

The cycle $\cC_L$ has only one equilibrium, $\xi_2$, where a transverse
eigenvalue of linearisation is positive. Therefore, theorem \ref{th_4} implies
that the cycle is f.a.s. whenever the matrix $M_L=M_{123}M_{412}M_{341}M_{234}$
satisfies the conditions (i)-(iii).
Multiplying the matrices (\ref{m14}) we obtain that
\begin{equation}\label{maA}
M_L=\left(
\begin{array}{cccc}
a_{11}&a_{12}&0&0\\
a_{21}&a_{22}&0&0\\
a_{31}&a_{32}&1&0\\
a_{41}&a_{42}&0&1
\end{array}
\right),
\end{equation}
where
\begin{equation}\label{bbb}
\renewcommand{\arraystretch}{1.2}
\begin{array}{l}
a_{11}=(b_{314}b_{421}+b_{324})(b_{132}b_{243}+b_{142})+b_{314}b_{431}b_{243}\\
a_{21}=(b_{314}b_{421}+b_{324})b_{132}b_{213}+b_{314}b_{431}b_{213}\\
a_{31}=(b_{314}b_{421}+b_{324})(b_{132}b_{253}+b_{152})+b_{314}b_{431}b_{253}\\
a_{41}=(b_{314}b_{421}+b_{324})(b_{132}b_{263}+b_{162})+b_{314}b_{431}b_{263}\\
a_{21}=b_{421}(b_{132}b_{243}+b_{142})+b_{431}b_{243},\quad
a_{22}=b_{421}b_{132}b_{213}+b_{431}b_{213}\\
a_{32}=b_{421}(b_{132}b_{253}+b_{152})+b_{431}b_{253},\quad
a_{42}=b_{421}(b_{132}b_{263}+b_{162})+b_{431}b_{263}.
\end{array}
\end{equation}
Due to the assumption that the network $\cX$ is stable in the transverse
directions, in (\ref{bbb}) the ratios $b_{ijk}$ are positive, except for
$b_{253}$. Therefore, the entries of matrix (\ref{maA}) are positive, except
possibly for $a_{31}$ and $a_{32}$.

The dominant eigenvalue of $M_L$, $\lambda_{\max}$, is the largest in absolute
value eigenvalue of its upper left $2\times2$ submatrix which we denote by
$M'_L$. The entries of $M'_L$ are positive, implying that
the discriminant of its characteristic polynomial $F(\lambda)$ is positive.
Hence, the eigenvalues of $M'_L$ are real, which implies that $\lambda_{\max}$
satisfies condition (i) of theorem \ref{th_4}. Since
$F(a_{11})=F(a_{22})=-a_{12}a_{21}<0$, the dominant eigenvalue satisfies
$\lambda_{\max}>\max(a_{11},a_{22})>0$. Condition (ii) is satisfied
whenever \cite{pa11}
\begin{equation}\label{cond1}
\max\biggl({a_{11}+a_{22}\over2},
a_{11}+a_{22}-a_{11}a_{22}+a_{12}a_{21}\biggr)>1.
\end{equation}
Let $\bv=(v_1,v_2,v_3,v_4)$ be the eigenvector of $A$ associated
with $\lambda_{\max}$ and $v_1=1$. The component $v_2$ is positive because
\begin{equation}\label{cond21}
v_2=a_{21}v_1/(\lambda_{\max}-a_{22}).
\end{equation}
From (\ref{maA}) the latter two components satisfy
\begin{equation}\label{cond22}
v_3={a_{31}v_1+a_{32}v_2\over\lambda_{\max}-1},\quad
v_4={a_{41}v_1+a_{42}v_2\over\lambda_{\max}-1}.
\end{equation}
Since $a_{41}>0$ and $a_{42}>0$, the latter equality implies that $v_4>0$.
Substituting (\ref{cond21}) into the first inequality in (\ref{cond22}) we
obtain that (iii) is equivalent to
\begin{equation}\label{cond2}
a_{31}(\lambda_{\max}-a_{22})+a_{32}a_{21}>0.
\end{equation}

Therefore, the conditions for fragmentary asymptotic stability of the cycle
$\cC_L$ are (\ref{cond1}) and (\ref{cond2}), where $\lambda_{\max}$ is
the largest eigenvalue of the upper left $2\times2$ submatrix of (\ref{maA})
and the dependence of its entries on $\mu_{ij}$ is given by (\ref{coeB}) and
(\ref{bbb}). The permutation of coordinates
$(x_1,x_2,x_3,x_4,x_5,x_6)\to(x_1,x_2,x_5,x_6,x_3,x_4)$ maps $\cC_L$ to
$\cC_R$. Hence, the condition for stability of $\cC_R$ are obtained from
those for $\cC_L$ by the permutation of subscripts $(123456)\to(125634)$
of $\mu_{ij}$ in (\ref{coeB}).

\subsection{Trail-stability of $\cC_{LR}$.}\label{sec32}

In this subsection we show that the omnicycle $\cC_{LR}$ can be
trail-stable. Stability of the cycle depends on eigenvalues and eigenvectors
of its $4\times4$ transition matrix. In order to be able to satisfy the
conditions for stability we make two assumptions.
Namely, we assume that the equation (\ref{eq_ode6}) has an additional symmetry
\begin{equation}\label{sym}
(x_1,x_2,x_3,x_4,x_5,x_6)\to(x_1,x_2,x_5,x_6,x_3,x_4)
\end{equation}
and that the eigenvalues
\begin{equation}\label{eig}
\mu_{24},\mu_{26},\mu_{31},\mu_{35},\mu_{36},
\mu_{42},\mu_{45},\mu_{46},
\mu_{52},\mu_{53},\mu_{54},\mu_{61},\mu_{63},\mu_{64}
\end{equation}
are much smaller than other eigenvalues of the Jacobians $\rd f(\xi_j)$.
To prove existence of a system where $\cC_{LR}$ is trail-stable, we first prove
that for vanishing eigenvalues (\ref{eig}) it is possible to identify other
eigenvalues $\mu_{ij}$ such that the transition matrix satisfies
conditions (i)-(iii) of theorem \ref{th_4}. The conditions are not strict and
eigenvalues and eigenvectors of a matrix depend continuously on its entries,
therefore (i)-(iii) remain true for sufficiently small eigenvalues in
(\ref{eig}) as well. For a given set of eigenvalues of $\rd f(\xi_j)$, a system
that has such eigenvalues can be constructed by the procedure discussed in
subsection \ref{sec51}.

Existence of the symmetry (\ref{sym}) implies that the omnicycle is comprised
of two building blocks (see definition \ref{bb}), therefore to study its
trail-stability it is sufficient to consider stability of the return map
$H^{out}_{23}\to H^{out}_{25}$
with respective transition matrix $M_{LR}=M_{125}M_{412}M_{341}M_{234}$.
Let the bases in $H^{out}_{12}$, $H^{out}_{34}$ and $H^{out}_{41}$ be given by
(\ref{bases}) and the basis in $H^{out}_{25}$ be $\{\be_6,\be_1,\be_3,\be_4\}$.
The basic transition matrices of the cycle $\cC_{LR}$ in a system with
the symmetry (\ref{sym}) and vanishing eigenvalues (\ref{eig}) are
\begin{equation}\label{mLR}
\renewcommand{\arraystretch}{1.7}
\begin{array}{cc}
M_{412}=\left(
\begin{array}{cccc}
b_{132}&1&0&0\\
b_{142}&0&0&0\\
b_{152}&0&1&0\\
b_{162}&0&0&1
\end{array}
\right),
&
M_{234}:=\left(
\begin{array}{cccc}
0&1&0&0\\
b_{324}&0&0&0\\
0&0&1&0\\
0&0&0&1
\end{array}
\right)\\
M_{341}=\left(
\begin{array}{cccc}
0&1&0&0\\
b_{431}&0&0&0\\
0&0&1&0\\
0&0&0&1
\end{array}
\right),
&
M_{125}=\left(
\begin{array}{cccc}
0&0&0&1\\
0&0&b_{215}&0\\
1&0&-1&0\\
0&1&0&0
\end{array}
\right)
\end{array}
\end{equation}
where $b_{ijk}$ are defined in (\ref{coeB}).

Multiplication of the matrices implies that
\begin{equation}\label{mLR1}
M_{LR}=\left(
\begin{array}{cccc}
\tilde a_{11}&0&0&1\\
\tilde a_{21}&0&\tilde a_{23}&0\\
0&\tilde a_{32}&-1&0\\
\tilde a_{41}&0&0&0
\end{array}
\right),
\end{equation}
where
\begin{equation}\label{mLR2}
\tilde a_{11}=b_{162}b_{324},\ \tilde a_{21}=b_{152}b_{324}b_{213},\
\tilde a_{23}=b_{213},\ \tilde a_{32}=b_{413},\ \tilde a_{41}=b_{142}b_{324}.
\end{equation}

Upon the permutation of bases in $H^{out}_{25}$:
$\{\be_6,\be_1,\be_3,\be_4\}\to\{\be_6,\be_4,\be_1,\be_3\}$, matrix
(\ref{mLR1}) takes the form
\begin{equation}\label{mLR3}
M_{LR}=\left(
\begin{array}{cccc}
\tilde a_{11}&1&0&0\\
\tilde a_{41}&0&0&0\\
\tilde a_{21}&0&0&\tilde a_{23}\\
0&0&\tilde a_{32}&\tilde a_{33}
\end{array}
\right).
\end{equation}
The eigenvalues of this matrix are
\begin{equation}\label{mLR4}
\lambda_{1,2}={\tilde a_{11}\pm(\tilde a_{11}^2+4\tilde a_{11})^{1/2}\over2},
\quad\lambda_{3,4}={-1\pm(1+4\tilde a_{23}\tilde a_{32})^{1/2}\over2},
\end{equation}
where the indices 1 and 3 correspond to positive signs in front of the square
roots. Therefore, the conditions (i) and (ii) of theorem \ref{th_4} for the
matrix $M_{LR}$ take the form
\begin{equation}\label{mLR5}
\lambda_1>\max(1,|\lambda_4|).
\end{equation}
From (\ref{mLR3}) the components of the eigenvector $\bv=(v_1,v_2,v_3,v_4)$
of $M_{LR}$ associated with $\lambda_1$ satisfy
$$v_2={v_1\tilde a_{41}\over\lambda_1},\ v_3={v_4\tilde a_{23}\over\lambda_1},\
v_4=v_1\tilde a_{31}
\biggl(\lambda_1+1-{\tilde a_{32}\tilde a_{23}\over\lambda_1}\biggr)^{-1}.$$
Therefore, the condition (iii) of theorem \ref{th_4} holds true whenever
\begin{equation}\label{mLR6}
(\lambda_1+1)\lambda_1-\tilde a_{32}\tilde a_{23}>0.
\end{equation}

Finally, we show that conditions (\ref{mLR5}),(\ref{mLR6}) imply that the
cycle $\cC_L$ is completely unstable. (Therefore, by the same arguments as
applied to show trail-stability of $\cC_{LR}$, it remains unstable for a
slightly perturbed non-vanishing eigenvalues listed in (\ref{eig}).)
The transition matrix (\ref{maA}) under the assumptions
of symmetry (\ref{sym}) and for vanishing eigenvalues (\ref{eig}) takes the form
\begin{equation}\label{maAA}
M_L=\left(
\begin{array}{cccc}
a_{11}&0&0&0\\
a_{21}&a_{22}&0&0\\
0&a_{32}&1&0\\
a_{41}&0&0&1
\end{array}
\right),
\end{equation}
where $a_{32}=b_{235}b_{413}<0$. The components of the dominant eigenvector
satisfy $v_3=v_1a_{32}/(\lambda_{\max}-1)$. Hence,
$\lambda_{\max}>1$ implies that $v_1v_3<0$, i.e., for the matrix $M_L$
the conditions (ii) and (iii) of theorem \ref{th_4} can not be satisfied
simultaneously.

\begin{remark}\label{rem1}
To show that the cycle $\cC_L$ can be stable in symmetric system (\ref{sym})
assume that the eigenvalues
\begin{equation}\label{list}
\mu_{13},\ \mu_{16},\ \mu_{24},\ \mu_{26},\ \mu_{3j},\ j=1,3,5;\
\mu_{41},\ \mu_{46}
\end{equation}
are much smaller that the other ones. Under the assumption of vanishing
eigenvalues (\ref{list}) the matrix (\ref{maA}) takes the form
\begin{equation}\label{maAAA}
M_L=\left(
\begin{array}{cccc}
b_{324}b_{142}&0&0&0\\
0&b_{213}b_{431}&0&0\\
0&b_{451}-b_{431}&1&0\\
b_{324}b_{142}&0&0&1
\end{array}
\right).
\end{equation}
The conditions (i)-(iii) are satisfied if
$$\max(b_{324}b_{142},b_{213}b_{431})>1\hbox{ and }b_{451}-b_{431}>0.$$
These conditions are also satisfied for sufficiently small eigenvalues
(\ref{list}).
\end{remark}

\section{Numerical examples}
\label{sec5}

In this section we study numerically behaviour of trajectories near the
heteroclinic networks $\cX=\cC_L\cup\cC_R$ shown in fig.~\ref{fig1}.
In subsection \ref{sec51} we discuss construction of a $\Z^6_2$-equivariant
dynamical system, possessing such a network with prescribed eigenvalues
of Jacobians $\rd f(\xi_j)$, $1\le j\le 6$. Then, using results of section
\ref{sec4}, we give numerical examples of attracting $\cC_L$ and $\cC_{LR}$.
Our simulations indicate that when none of $\cC_L$, $\cC_R$ or $\cC_{LR}$
is attracting, nearby trajectories nevertheless can be attracted to the network
displaying regular or irregular switching between $\cC_L$ and $\cC_R$.

\subsection{Construction of the system}\label{sec51}

Consider the system
\begin{equation}\label{sysf}
\dot x_i=x_i(-1+\sum_{1\le j\le 6} \beta_{ij}x_j^2),\quad
\hbox{ where }\beta_{ii}=1,\ 0\le i\le 6.
\end{equation}
By construction (\ref{sysf}) is $\Z^6_2$-equivariant with
the symmetry group generated by the inversions of Cartesian coordinates.
Any coordinate axis, plane or a hyperplane is an invariant subspace of
(\ref{sysf}). It has equilibria $\xi_i=\pm1$ on each of the coordinate axes,
which are attractive in the radial direction.

The restriction of (\ref{sysf}) to the coordinate plane $<\be_l,\be_k>$ is
\begin{equation}\label{syskl}
\dot x_l=x_l(-1+x_l^2+\beta_{lk}x_k^2),\ \dot x_k=x_k(-1+x_k^2+\beta_{kl}x_l^2).
\end{equation}
It is known (see, e.g., \cite{op09} and references therein) that if both
$\xi_l$ or $\xi_k$ are stable in the radial direction, one of them is unstable
in the orthogonal direction and the other equilibrium is stable, then there
exists a robust heteroclinic trajectory connecting these equilibria.

The eigenvalues of $\rd f(\xi_l)$ and $\rd f(\xi_k)$ in the directions of
$\be_k$ and $\be_k$, respectively, are
\begin{equation}\label{lambb}
\mu_{lk}=-1+\beta_{kl},\ \mu_{kl}=-1+\beta_{lk}.
\end{equation}
Each of $\beta_{ij}$ enters into only one expression for $\mu_{kl}$, therefore
we can design system (\ref{sysf}) for any prescribed $\mu_{kl}$ compatibles
with the connections shown in fig.~\ref{fig1}. (Except for radial eigenvalues, which
do not enter into the expressions for stability. These eigenvalues can be
altered by modifying $b_{ii}$.)

\subsection{Numerical results: symmetric system}\label{sec42}

\begin{figure}
{\Large
\hspace*{-12mm}\includegraphics[width=18cm]{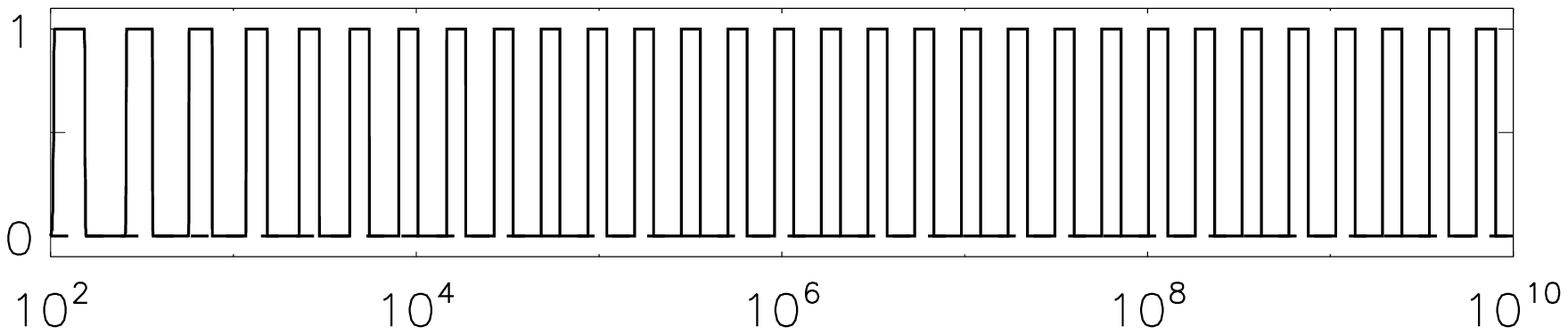}

\vspace*{-18mm}\hspace*{138mm}$\tau$

\vspace*{-5mm}\hspace*{65mm}(a)

\vspace{-4mm}

\hspace*{-12mm}\includegraphics[width=18cm]{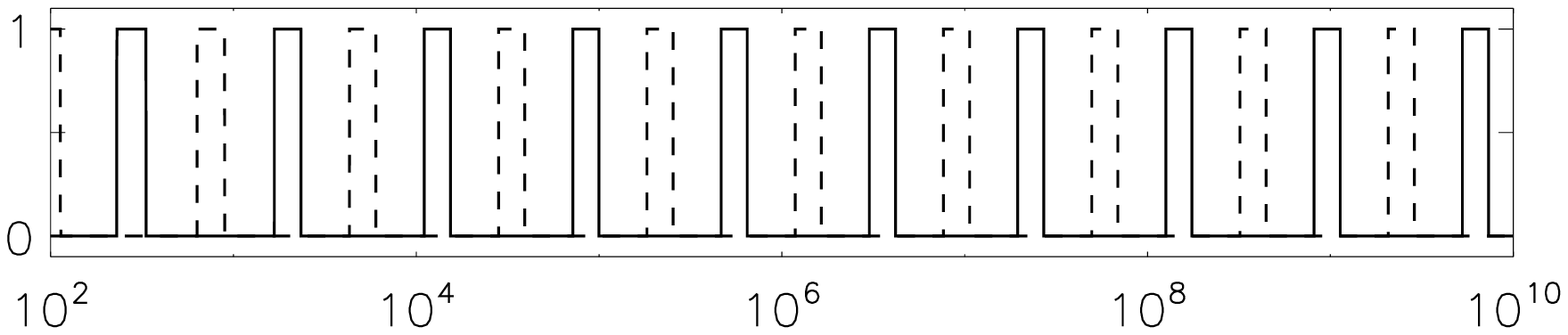}

\vspace*{-18mm}\hspace*{138mm}$\tau$

\vspace*{-5mm}\hspace*{65mm}(b)

\vspace{-4mm}

\hspace*{-12mm}\includegraphics[width=18cm]{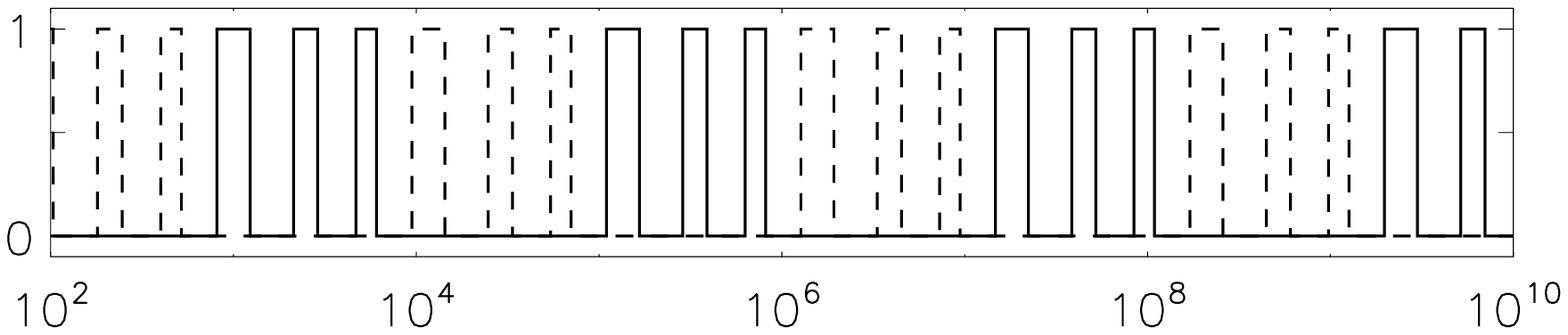}

\vspace*{-18mm}\hspace*{138mm}$\tau$

\vspace*{-5mm}\hspace*{65mm}(c)

\vspace{-4mm}
\hspace*{-12mm}\includegraphics[width=18cm]{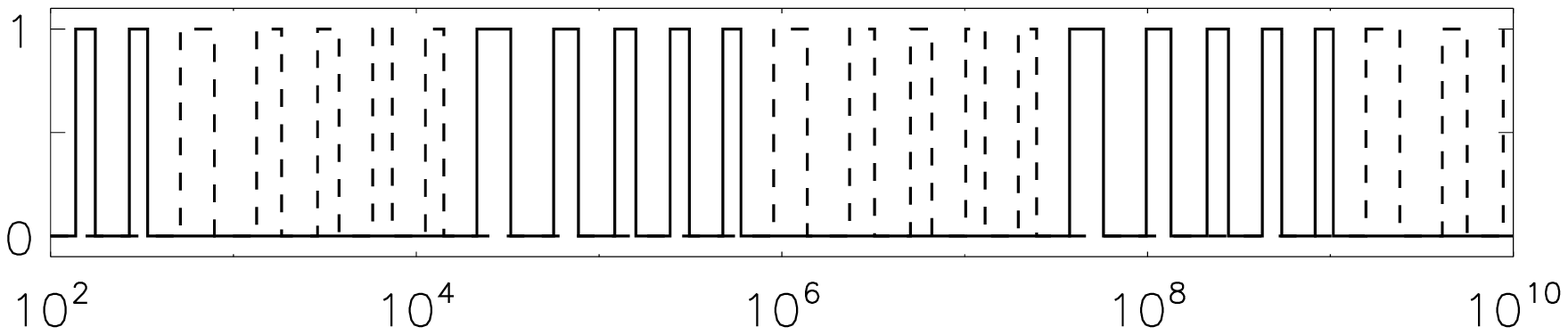}

\vspace*{-18mm}\hspace*{138mm}$\tau$

\vspace*{-5mm}\hspace*{65mm}(d)

\vspace{-4mm}
\hspace*{-12mm}\includegraphics[width=18cm]{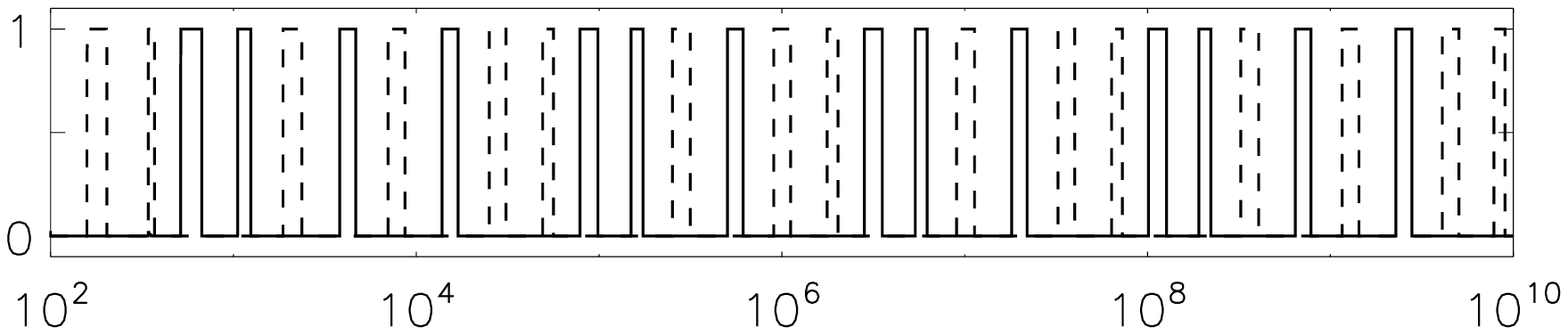}

\vspace*{-18mm}\hspace*{138mm}$\tau$

\vspace*{-5mm}\hspace*{65mm}(e)

\vspace{-2mm}
}

\caption{The dependence on time of $x_3$ (solid line) and
$x_5$ (dashed line).
\label{fig2}}
\end{figure}

Here we present results of numerical simulations (see fig.~\ref{fig2}) of
solutions to equations (\ref{sysf}) in the case when the system has additional
symmetry (\ref{sym}).
In section \ref{sec3} we derived conditions for stability of heteroclinic cycles
$\cC_L$ and $\cC_R$, and omnicycle $\cC_{LR}$, which involve ratios
$b_{ijk}=-\mu_{ij}/\mu_{jk}$. Therefore, we can choose $\mu_{lk}$ to make
the cycles is stable or unstable. The values of $\mu_{lk}$ employed in
simulations shown in the figure are given in table~\ref{tab1}.

\begin{table}[ht]
%\hskip -1cm
\begin{tabular}{|c|c|}
\hline plate & eigenvalues \\ \hline
(a) & $\mu_{12}=\mu_{34}=\mu_{41}=\mu_{56}=\mu_{61}=1,\ \mu_{14}=\mu_{16}=-1.2$\\
& $\mu_{21}=-1,\ \mu_{23}=\mu_{25}=1/1.4,\ \mu_{32}=\mu_{52}=-1.5,\
\mu_{43}=\mu_{65}=-1.1,\ \mu_{45}=\mu_{63}=-1.7$\\\hline
(b) & $\mu_{12}=\mu_{34}=\mu_{41}=\mu_{56}=\mu_{61}=1,\ \mu_{14}=\mu_{16}=-1.2$\\
& $\mu_{21}=-1,\ \mu_{23}=\mu_{25}=1/1.4,\ \mu_{32}=\mu_{52}=-1.5,\
\mu_{35}=\mu_{53}=-0.4,\ \mu_{43}=\mu_{65}=-1.1$\\\hline
(c) & $\mu_{12}=\mu_{34}=\mu_{41}=\mu_{56}=\mu_{61}=1,\ \mu_{14}=\mu_{16}=-1.2$\\
& $\mu_{21}=-1,\ \mu_{23}=\mu_{25}=1/1.4,\ \mu_{32}=\mu_{52}=-1.5,\
\mu_{35}=\mu_{53}=-0.4,\ \mu_{43}=\mu_{65}=-1.6$\\\hline
(d) & $\mu_{12}=\mu_{34}=\mu_{41}=\mu_{56}=\mu_{61}=1,\ \mu_{14}=\mu_{16}=-1.2$\\
& $\mu_{21}=-1,\ \mu_{23}=\mu_{25}=1/1.4,\ \mu_{32}=\mu_{52}=-1.5,\
\mu_{35}=\mu_{53}=-0.4,\ \mu_{43}=\mu_{65}=-1.4$\\\hline
(e) & $\mu_{12}=\mu_{34}=\mu_{41}=\mu_{56}=\mu_{61}=1,\ \mu_{14}=\mu_{16}=-1.2$\\
& $\mu_{21}=-1,\ \mu_{23}=\mu_{25}=1/1.4,\ \mu_{32}=\mu_{52}=-1,\
\mu_{15}=\mu_{13}=-0.4,\ \mu_{43}=\mu_{65}=-2$\\\hline
\end{tabular}
\caption{Eigenvalues of linearisations employed in numerical simulations
shown in fig.~\ref{fig2}. We do not list $\mu_{ii}=-1$ or eigenvalues that
are equal to -0.01.}\label{tab1}
\end{table}

We start from stable $\cC_L$ and $\cC_{LR}$ (plates (a) and (b)\,) where the
values of $\mu_{lk}$ given in table 1 are taken to satisfy conditions
(\ref{cond1}),(\ref{cond2}) and (\ref{mLR2}),(\ref{mLR4})-(\ref{mLR6}).
For all other considered variants, where $\mu_{lk}$ are taken so that both
$\cC_L$ and $\cC_{LR}$ are unstable, the trajectories staying near the network
display regular or irregular behaviour: they can switch between $\cC_L$ and
$\cC_R$ making a fixed number of turns (this number depends on parameters of
(\ref{sysf})\,) around one cycle before switching to the other one, or this
number can vary. Examples of such regular of irregular switching are shown in
plates (c)-(e).

For $\tau\to\infty$ the trajectories are attracted by the heteroclinic network,
which is indicated by exponentially increasing subsequent time intervals
between visiting $\xi_2$
\footnote{On the figures only the coordinates $x_3$ and $x_5$ are displayed:
$x_3$ close to one indicates that after $\xi_2$ a trajectory goes to $\cC_L$,
while $x_5$ close to one corresponds to the trajectory going to $\cC_R$.}
(i.e., time intervals needed to make a turn around
$\cC_L$ or $\cC_R$). On the plots the horisontal axis (time) is scales
exponentially, implying that temporal behaviour is visually similar to periodic.

\begin{remark}\label{rem00}
In simulations the values of $\beta_{lk}$ are taken such that the eigenvalues of
transition matrices $M_L$ and $M_R$ satisfy $\lambda_L^{\max}>1$
and $\lambda_R^{\max}>1$. We can not prove that these inequalities imply
fragmentary asymptotic stability of the network, however numerical results
indicate that this might be the case, because for all performed runs
trajectories were attracted by the network.
\end{remark}

\subsection{Numerical results: non-symmetric system}\label{sec52}

\begin{figure}
{\Large
\hspace*{-12mm}\includegraphics[width=18cm]{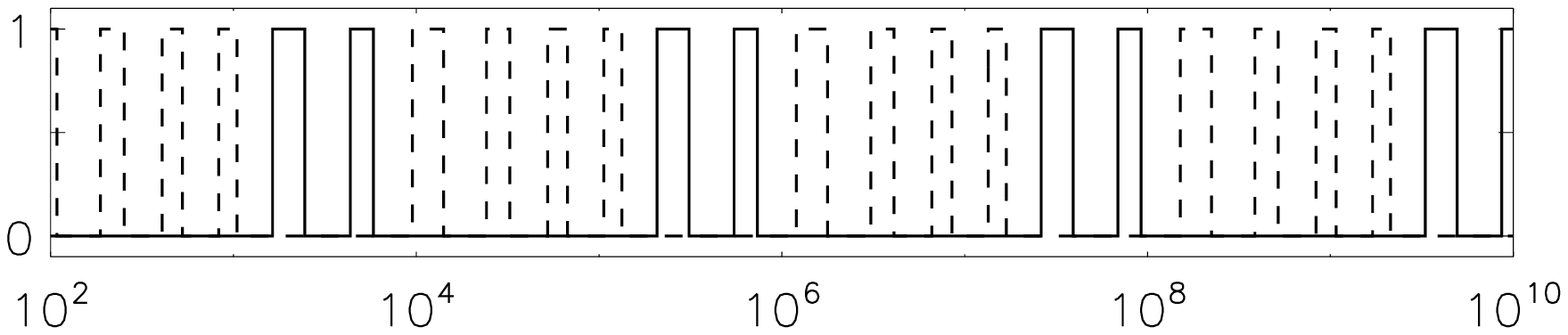}

\vspace*{-18mm}\hspace*{138mm}$\tau$

\vspace*{-5mm}\hspace*{65mm}(a)

\vspace{-4mm}

\hspace*{-12mm}\includegraphics[width=18cm]{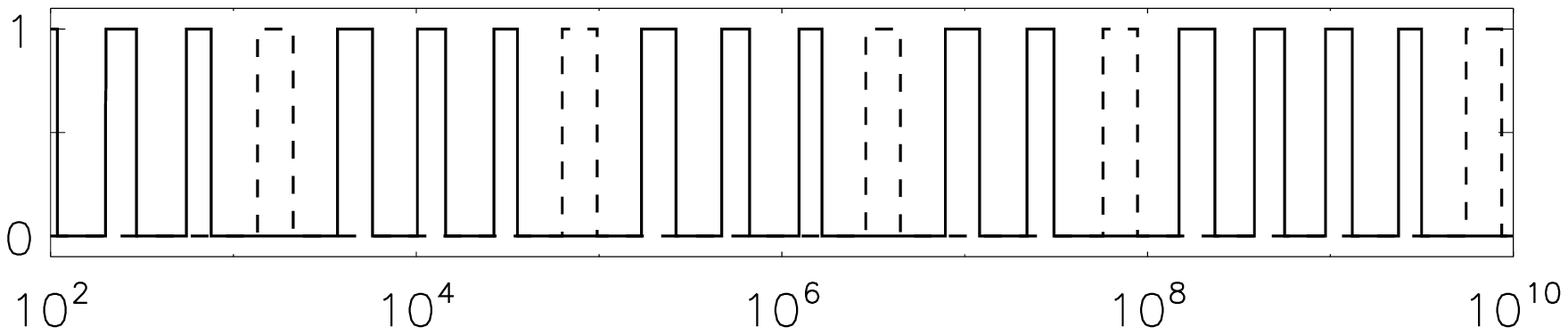}

\vspace*{-18mm}\hspace*{138mm}$\tau$

\vspace*{-5mm}\hspace*{65mm}(b)

\vspace{-4mm}

\hspace*{-12mm}\includegraphics[width=18cm]{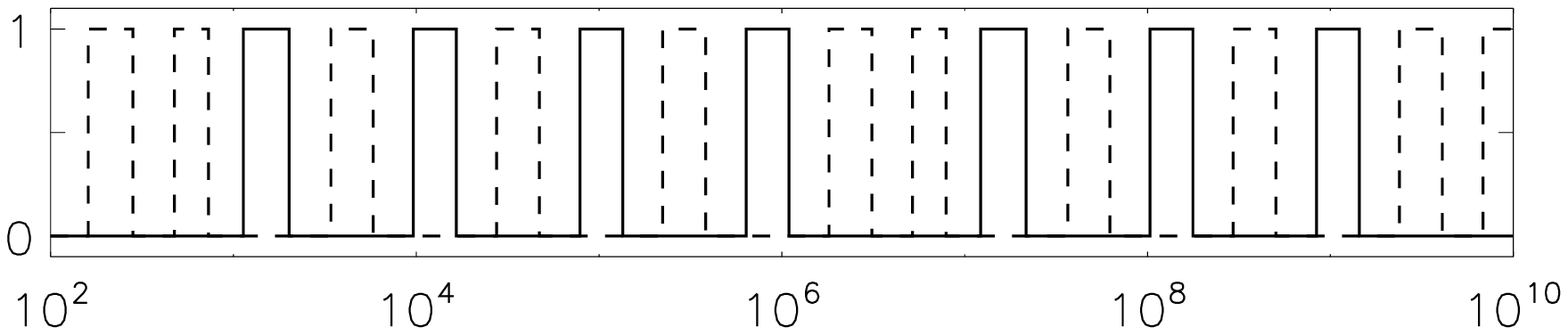}

\vspace*{-18mm}\hspace*{138mm}$\tau$

\vspace*{-5mm}\hspace*{65mm}(c)

\vspace{-4mm}
\hspace*{-12mm}\includegraphics[width=18cm]{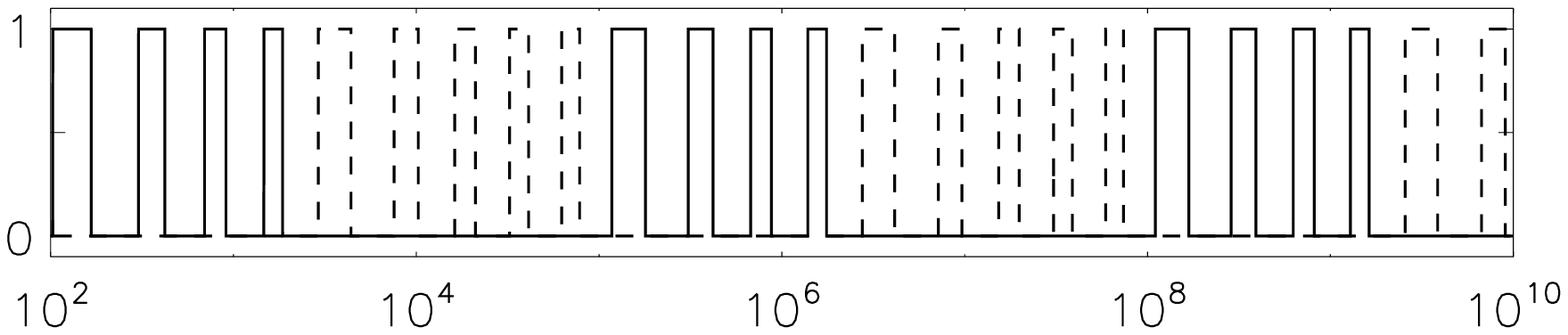}

\vspace*{-18mm}\hspace*{138mm}$\tau$

\vspace*{-5mm}\hspace*{65mm}(d)

\vspace{-4mm}
\hspace*{-12mm}\includegraphics[width=18cm]{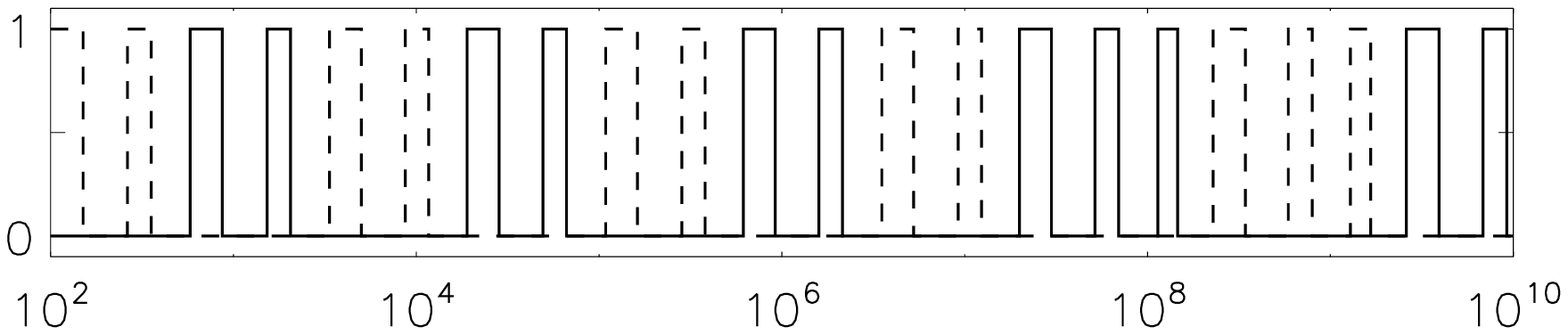}

\vspace*{-18mm}\hspace*{138mm}$\tau$

\vspace*{-5mm}\hspace*{65mm}(e)

\vspace{-2mm}
}

\caption{The dependence on time of $x_3$ (solid line) and
$x_5$ (dashed line).
\label{fig3}}
\end{figure}

In this section we present results of numerical integration of system
(\ref{sysf}) when additional symmetry (\ref{sym}) is not imposed. Typical
behaviour of trajectories near the network is shown in fig.\ref{fig3} and the
respective eigenvalues of $\rd f(\xi_j)$ are given in table \ref{tab2}.
The eigenvalues are chosen such that none of $\cC_L$ or $\cC_R$ is
f.a.s.

\begin{table}[ht]
%\hskip -1cm
\begin{tabular}{|c|c|}
\hline plate & eigenvalues \\ \hline

(a) & $\mu_{12}=\mu_{34}=\mu_{41}=\mu_{56}=\mu_{61}=1,\ \mu_{14}=\mu_{16}=-1.2,\ \mu_{21}=-1,\ \mu_{23}=1/1.4$\\
& $\mu_{25}=1/1.5,\ \mu_{32}=\mu_{52}=-1.5,\
\mu_{43}=1.8,\ \mu_{65}=-1.4,\ \mu_{35}=-0.4,\ \mu_{53}=-0.5$\\\hline
(b) & $\mu_{12}=\mu_{34}=\mu_{41}=\mu_{56}=\mu_{61}=1,\ \mu_{14}=\mu_{16}=-1.4,\ \mu_{21}=-1$\\
& $\mu_{23}=\mu_{25}=1/1.4,\ \mu_{32}=\mu_{52}=-1.5,\
\mu_{15}=-0.4,\ \mu_{43}=2.9,\ \mu_{65}=-1.6$\\\hline
(c) & $\mu_{12}=\mu_{34}=\mu_{41}=\mu_{56}=\mu_{61}=1,\ \mu_{14}=\mu_{16}=-1.4$\\
& $\mu_{21}=-1,\ \mu_{23}=\mu_{25}=1/1.4,\ \mu_{32}=\mu_{52}=-1.5,\
\mu_{43}=2.9,\ \mu_{65}=-1.6$\\\hline
(d) & $\mu_{12}=\mu_{34}=\mu_{41}=\mu_{56}=\mu_{61}=1,\ \mu_{14}=\mu_{16}=-1.2,\ \mu_{21}=-1,\ \mu_{23}=1/1.4$\\
& $\mu_{25}=1/1.7,\ \mu_{32}=\mu_{52}=-1.5,\
\mu_{35}=\mu_{53}=-0.4,\ \mu_{43}=-1.6,\ \mu_{65}=-1.1$\\\hline
(e) & $\mu_{12}=\mu_{34}=\mu_{41}=\mu_{56}=\mu_{61}=1,\ \mu_{14}=\mu_{16}=-1.2,\ \mu_{21}=-1,\ \mu_{23}=1/1.4$\\
& $\mu_{25}=1/1.5,\ \mu_{32}=\mu_{52}=-1.5,\
\mu_{35}=\mu_{53}=-0.4,\ \mu_{43}=-1.7,\ \mu_{65}=-1.5$\\\hline
\end{tabular}
\caption{Eigenvalues of linearisations employed in numerical simulations
shown in fig.~\ref{fig3}. We do not list $\mu_{ii}=-1$ or eigenvalues that
are equal to -0.01.}\label{tab2}
\end{table}

A regular behaviour (examples are shown in plates (a) and (d)\,) can be
described as a repeating
pattern comprised of $n_L$ iterations along $\cC_L$ followed by $n_R$
iterations along $\cC_R$. Apparently, this indicates that the respective
omnicycle, which can be labelled as $(n_L,n_R)$, is f.a.s.
Often we observe less regular
behaviour (as shown in plates (b), (c) and (e)\,) which either indicate chaos
or a long omnicycle.
Due to the presence of numerical errors, it makes no sense to investigate
which kind of behaviour, chaotic or periodic, takes place: what we see in
simulations may be different from what really takes place in the system.

\section{Conclusion}\label{sec6}

In this paper we studied behaviour of trajectories near a network in
$\R^6$ comprised of two heteroclinic cycles. We derived condition for f.a.s.
of these heteroclinic cycles, proved that the network can be f.a.s. while
both cycles are completely unstable and presented results of numerical simulations
indicating that for a positive measure set of initial conditions the behaviour
of nearby trajectories is apparently chaotic.

To prove stability of the network we introduce the notions of an omnnicycle
and trail-stability, which were implicitly used in earlier studies of behaviour
of trajectories near a heteroclinic network. The definition of
omnicycle is similar to the one of heteroclinic cycle, except that the equilibria
and heteroclinic connections are not required to be distinct. Hence, the
conditions for trail-stability of an omnicycle are identical to the condition
of f.a.s. of a heteroclinic cycle, for type Z objects they
were proven in \cite{op12}.

To prove possibility of the existence of an attracting omnicycle we have assumed
that some
eigenvalues of linearisations are much smaller than the others, which allows
us to calculate eigenvalues and eigenvectors of transition matrices.
The assumption of smallness of some of the eigenvalues of linearisations
can be useful in other studies of stability of cycles, omnicycles or
heteroclinic networks in high-dimensional spaces, where direct calculation
of eigenvalues is not possible due to large dimensions of transition matrices.

%\bigskip\noindent
%Data availability\\
%The data that support the findings of this study are available within the
%article.
\subsection*{Acknowledgements}

I am grateful to Alexander Lohse for a discussion of definition \ref{defn1}.

\end{document}